\documentclass[journal=jacsat,manuscript=article]{achemso}
\setkeys{acs}{maxauthors=100}

\SectionsOn
\SectionNumbersOn
\usepackage[T1]{fontenc} 
\usepackage[utf8]{inputenc}
\usepackage{amsmath,amssymb}
\usepackage{multirow}
\usepackage{comment}
\usepackage{booktabs}
\usepackage{geometry}
\usepackage{filecontents}

\DeclareUnicodeCharacter{2005}{}
\newcommand{\name}{SPARROW}

\newcommand{\reward}{U}
\newcommand{\mollegend}{Molecules in blue are starting materials, and those in pink are targets. Reaction conditions and scores were obtained using ASKCOS \cite{gao_using_2018, coley_robotic_2019, coley_graph-convolutional_2019}. Compound costs were sourced from the ChemSpace API \cite{noauthor_chemspace_nodate}.}

\usepackage{xr}
\makeatletter
\newcommand*{\addFileDependency}[1]{
\typeout{(#1)}
\@addtofilelist{#1}
\IfFileExists{#1}{}{\typeout{No file #1.}}
}\makeatother
\addFileDependency{supplementary_v2.tex}%
\addFileDependency{supplementary_v2.aux}%

\author{Jenna C. Fromer}
\author{Connor W. Coley}
\email{ccoley@mit.edu}
\affiliation[Unknown University]
{Department of Chemical Engineering, MIT, Cambridge, MA 02139}
\alsoaffiliation[Second University]
{Department of Electrical Engineering and Computer Science, MIT, Cambridge, MA
02139}

\title[]
  {An algorithmic framework for synthetic cost-aware decision making in molecular design }

\abbreviations{}
\keywords{American Chemical Society, \LaTeX}

\begin{document}

\begin{abstract}
Small molecules exhibiting desirable property profiles are often discovered through an iterative process of designing, synthesizing, and testing sets of molecules. The selection of molecules to synthesize from all possible candidates is a complex decision-making process that typically relies on expert chemist intuition. We propose a quantitative decision-making framework, SPARROW, that prioritizes molecules for evaluation by balancing expected information gain and synthetic cost. SPARROW integrates molecular design, property prediction, and retrosynthetic planning to balance the utility of testing a molecule with the cost of batch synthesis. We demonstrate through three case studies that the developed algorithm captures the non-additive costs inherent to batch synthesis, leverages common reaction steps and intermediates, and scales to hundreds of molecules. SPARROW is open source and can be found at \url{github.com/coleygroup/sparrow}.
\end{abstract}

\section*{Introduction}

Small molecules exhibiting desirable property profiles are often optimized through an iterative process of designing, synthesizing, and testing sets of molecules to elucidate the relationship between structure and function. The key challenge in each design iteration is to downselect and prioritize, among all possible molecules that \emph{could} be made and tested, which candidates are worth pursuing. This challenge exists across molecular design applications, including in the discovery of therapeutic candidates, organic catalysts, battery materials, and sustainable solvents. 

A myriad of computational workflows can aid in the prioritization of molecules, but they each make simplifying assumptions about the cost and utility associated with testing a given molecule. 
Generative models, for example, often propose molecules that are 
impractical to synthesize \cite{gao_synthesizability_2020, mendez-lucio_novo_2020} and therefore exceptionally costly to evaluate. 
Synthetic complexity or accessibility score filters \cite{ertl_estimation_2009, coley_scscore_2018, thakkar_retrosynthetic_2021, liu_retrognn_2022} may mitigate the burden of manually inspecting molecules for synthesizability \cite{andersson_making_2009}. However, they fail to capture the non-additive costs of synthesizing a batch of molecules. Specifically, filters that assess compounds individually cannot incorporate the cost savings associated with common intermediates and starting materials in batch synthesis. The same limitation applies to retrosynthetic software filters \cite{segler_planning_2018, coley_robotic_2019, genheden_aizynthfinder_2020, badowski_selection_2019} and reaction-based generative models\cite{gao_amortized_2022, zhang_bayesian_2023}.   
The consideration of synthetic cost may be better described as an art than a science at present, explaining the lack of quantitative decision-making frameworks that we feel are suitable for automatically selecting molecules, for example, in a lead optimization campaign \cite{breznik_prioritizing_2023}. 

The framework of Bayesian optimization \cite{frazier_tutorial_2018, shahriari_taking_2016, korovina_chembo_2020, pyzer-knapp_bayesian_2018} partially captures the balance of cost and experimental value that is inherent to iterative design cycles. 
However, even cost-aware Bayesian optimization approaches \cite{sasena_flexibility_2002, huang_sequential_2006, palizhati_agents_2022, zanjani_foumani_multi-fidelity_2023} presume a specific numerical cost for each experiment and cannot capture the non-additivity of synthetic costs for a batch of multiple molecules. The use of common intermediates and starting materials, parallel library chemistry, and laboratory automation can significantly influence the cost of molecular synthesis. Existing algorithms for the downselection of compounds and synthetic routes fail to grasp these effects. Methods that appropriately accommodate the value and cost of a hypothetical set of experiments could both accelerate molecular design campaigns and expand the adoption of computer-aided molecular design tools. 
 
\begin{figure}
    \centering
    \includegraphics{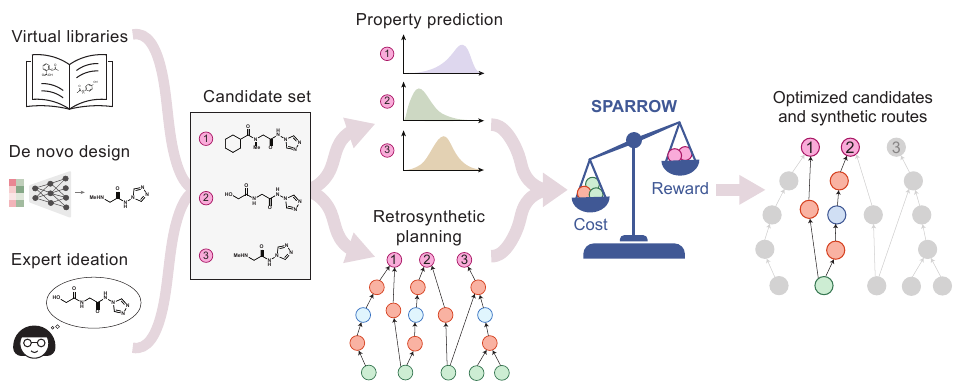}
    \caption{Overview of SPARROW and its role within the molecular design cycle. Each molecule in a candidate set, comprising molecular ideas from any combination of algorithmic or expert sources, is annotated with its anticipated properties and potential synthetic routes. These annotations can make use of quantitative structure-property relationship models with or without uncertainty quantification as well as computer-aided synthesis planning tools or human experts. SPARROW then weighs the utility of every candidate against their synthetic costs, not one-by-one but as a batch, and selects an optimal subset of candidates for synthesis and testing. }
    \label{fig:overview}
\end{figure}

In this work, we propose
Synthesis Planning And Rewards-based Route Optimization Workflow (SPARROW): an algorithmic decision-making framework for driving design cycles (Figure~\ref{fig:overview}). This work builds upon prior problem formulations for the simultaneous selection of synthetic routes to multiple molecules \cite{gao_direct_2021, gao_combining_2020, molga_computational_2019} and the integration of product and process systems design \cite{marvin_automated_2013, dahmen_model-based_2017, konig_integrated_2020, adjiman_process_2021}. Unlike alternative methods for downselection, SPARROW prioritizes molecules and their hypothetical synthetic routes from a pool of candidates using a multi-objective optimization criterion that balances cost and utility. The novelty of our contribution lies in the mathematical formulation of SPARROW's graph-based optimization; this allows SPARROW to accurately capture the expert intuition typically required for simultaneous molecule and synthetic route selection. SPARROW's impact will be further strengthened by advances in generative modeling for design ideation, computer-aided synthesis planning, and structure-property relationship modeling and uncertainty quantification. An open source implementation of SPARROW written in Python is made available at \url{github.com/coleygroup/sparrow}.

We demonstrate SPARROW's 
ability to orchestrate molecular design cycles through three case studies.
These applications illustrate how SPARROW 
(1) successfully balances information gain and synthetic cost, 
(2) captures the non-additivity of synthetic costs for a batch of molecules, 
and (3) scales to candidate libraries of hundreds of molecules. Importantly, 
SPARROW provides a unified framework for simultaneously evaluating molecule suggestions from virtual libraries, de novo design algorithms, and human experts as well as synthetic route suggestions from retrosynthesis software and human experts.

\section*{Results}

SPARROW generates a reaction network comprised of candidate target molecules and synthetic routes, discussed in detail in the Methods section. A graph-based optimization problem is solved to downselect a set of molecules and synthetic routes that optimally balance the cumulative synthetic cost and utility, as shown in Figure \ref{fig:overview}. In this context, utility measures the value of evaluating the properties of a molecule. An appropriate measure of utility will vary across applications and at different stages of design. It may incorporate molecular property predictions, uncertainty in those predictions, or the potential for a new datapoint to improve a structure-property relationship. A candidate library must be provided to SPARROW with corresponding \emph{rewards} that indicate the utility associated with each candidate molecule. 

\begin{figure}
    \centering
    \includegraphics{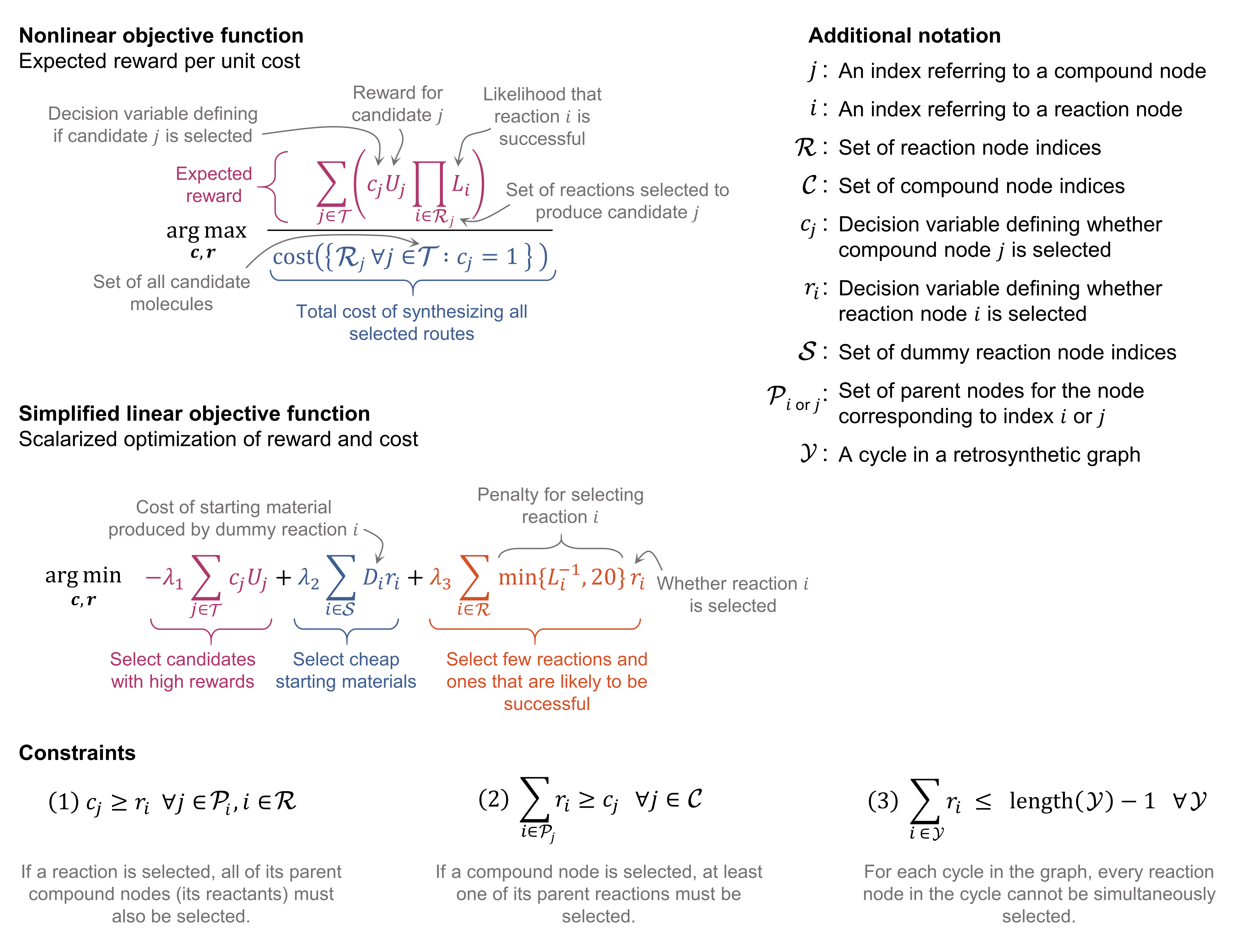}
    \caption{SPARROW's problem formulation. A nonlinear objective function is defined that maximizes the expected reward per unit cost of selected candidates and routes. We currently simplify this into a tractable objective function that balances utility and cost through a weighted sum. Three constraints are included to ensure that selected compounds have reactions to produce them, selected reactions have reactants to run them, and cycles are forbidden. }
    \label{fig:formulation}
\end{figure}

The reward achieved by selecting a molecule also depends on the success of the reaction steps selected to synthesize it. No information can be gained if a reaction step in the route to a candidate fails. We formalize this by aiming to maximize the \emph{expected} reward of selecting a candidate molecule, which can be represented by its reward multiplied by the probability that it is successfully synthesized (Figure \ref{fig:formulation}).  Balancing cost and utility, the objective of SPARROW may be formalized as the expected reward for all selected targets (i.e. candidates) divided by the cost of synthesizing all selected targets using selected routes. As described in the Methods section, we transform the expected reward per cost to a scalarized objective function that simultaneously maximizes cumulative utility and minimizes synthetic cost and risk of reaction failure. Three terms are optimized: (1) the cumulative rewards of selected candidates, (2) the cost of bought starting materials, and (3) the number of reactions and their likelihood of success, respectively. The linear optimization problem solved within SPARROW is defined mathematically in Figure \ref{fig:formulation}. Positive weighting factors ($\boldsymbol{\lambda} = [\lambda_1, \lambda_2, \lambda_3]$) assign relative importance to each objective and can be tuned to arrive at solutions that distinctly balance cost and utility. Higher values of $\lambda_1$ generally lead to solutions with more selected candidates and higher cumulative reward values. Higher values of $\lambda_2$ and $\lambda_3$ impose greater penalties for selecting costly starting materials and for selecting many reactions or ones with lower perceived likelihood of experimental success, respectively.

SPARROW was tested with three cases studies to showcase its ability to identify cost-efficient routes, balance information gain and cost, and unify library-based and de novo design. Our demonstrations illustrate that SPARROW can scale to the low hundreds of molecules. 

In all cases, we used ASKCOS \cite{coley_prediction_2017, coley_robotic_2019} with an expansion time of 60 seconds to generate retrosynthetic graphs and estimate the likelihood of reaction success. Compound buyability and cost, if applicable, were assigned with the ChemSpace API using availability and pricing as of October 2023 \cite{noauthor_chemspace_nodate}. For baseline solutions, compound prices were sourced from the ChemSpace API in March 2024. A comparison of compound prices between October 2023 and March 2024, shown in Figure \ref{S-fig:cost-comparison}, demonstrates that cost discrepancies are minor and do not artificially boost SPARROW's performance relative to baselines. 

\subsection*{Balancing cost and utility}
Our first demonstration of \name{} is on a candidate library of 14 molecules tested as inhibitors of the Alanine-Serine-Cysteine Transporter 2 (ASCT2) by \citet{garibsingh_rational_2021}. They selected molecules using molecular docking and binding free energy simulations prior to synthesis and testing. Although the molecules in this study were synthesized over multiple design cycles, the set of structures with binding free energy predictions serves as a representative case study for SPARROW. For candidate reward values, binding free energies for the full candidate set were scaled linearly between 0 and 1, with the most negative binding energy mapping to 1 and the most positive mapping to 0.

\newgeometry{top=1.5cm,bottom=1.5cm, left=1.5cm, right=1.5cm} 
\begin{figure}
    \centering
    \includegraphics{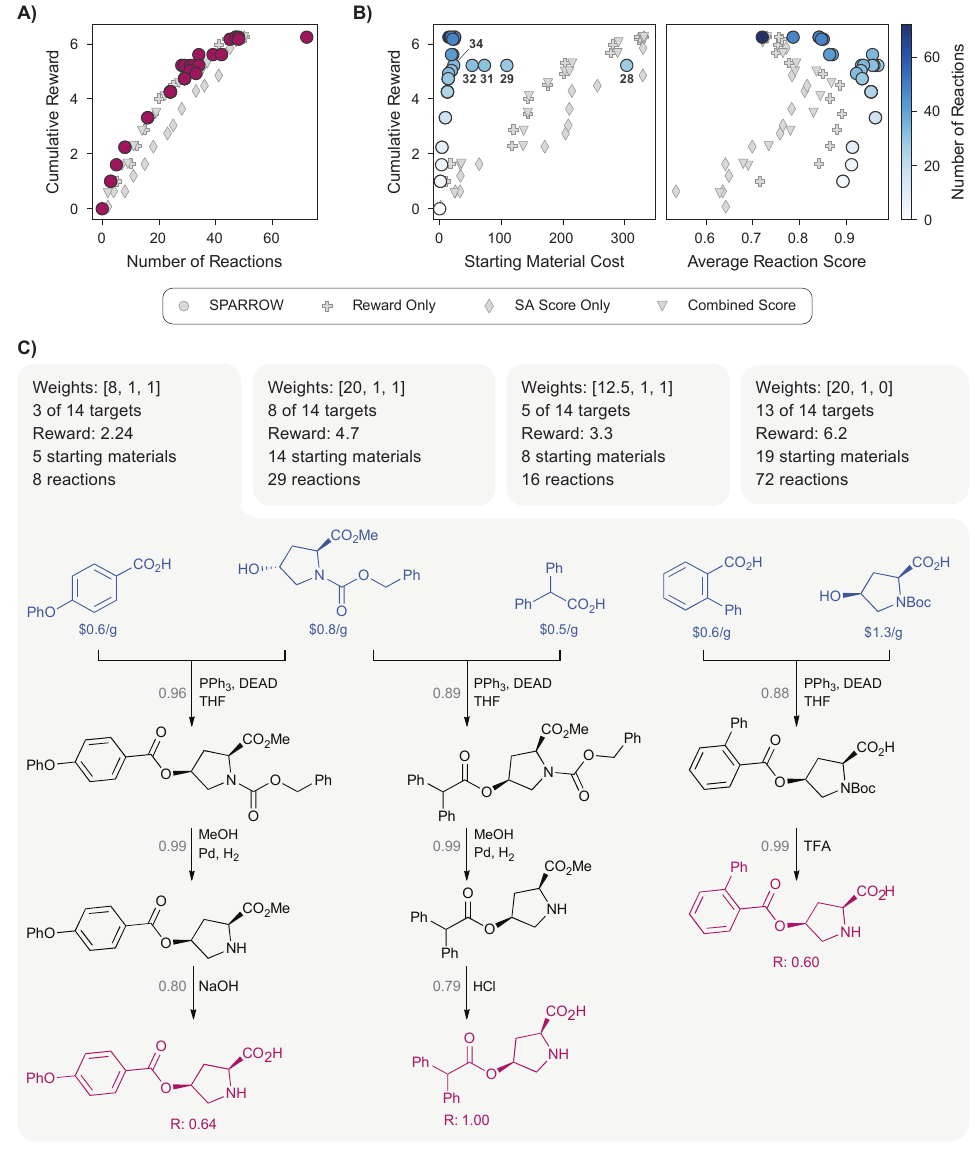}
    \caption{Demonstration of SPARROW's ability to balance cost and reward on a 14-member candidate library of putative ASCT2 inhibitors. (A, B) Weighting factors were varied to arrive at Pareto fronts that depict the trade-offs between cumulative reward, starting material cost, number of reactions, and reaction scores (Table \ref{S-tab:case-1}). SPARROW identifies solutions with cheaper starting materials and higher reaction scores when compared to baselines (Table \ref{S-tab:case-1-baselines}). Some points are labeled with the corresponding number of reactions. (C) Specific solutions for different weighting factors $\boldsymbol{\lambda}$ are summarized, with the complete set of selected routes for $\boldsymbol{\lambda} = [8, 1, 1]$ drawn. \mollegend ~Note that the reported starting material cost is the sum of the price per gram of selected starting materials without consideration of the amount needed. }
    \label{fig:case-1}
\end{figure}
\restoregeometry

SPARROW identifies various synthetic plans that distinctly balance information gain and cost (Figure \ref{fig:case-1}). The cumulative utility, number of reaction steps, and total starting material cost of the synthetic routes proposed by SPARROW are strongly dependent on the scalarization weights $\boldsymbol{\lambda}$ (Figure \ref{fig:case-1}). This allows a user's relative preferences for cost and information gain to directly impact SPARROW's solution. The sample routes shown in Figure \ref{fig:case-1}C reveal that SPARROW prioritizes molecules that both have high rewards and can be synthesized in few steps from cheap starting materials. When possible, SPARROW utilizes common starting materials and overlapping reaction steps to reduce the overall cost of synthesizing a batch of molecules.

 We compare SPARROW's performance with alternative downselection strategies that do not consider the non-additivity of batch synthetic costs: selection based on reward only, synthetic accessibility score \cite{ertl_estimation_2009} only, and a combined score. For each strategy, a variety of solutions may be obtained by varying the number of selected compounds. In this case, SPARROW and baseline solutions require a comparable number of reaction steps (Figure \ref{fig:case-1}A). However, the routes selected by SPARROW require cheaper starting materials and use reactions with higher scores when compared to baselines, indicating higher model confidence in the likelihood of successful synthesis (Figure \ref{fig:case-1}B).

\subsection*{Unifying library-based and de novo design}

A second case study exemplifies SPARROW's ability to leverage common intermediates and unify library-based and de novo design. \citet{koscher_autonomous_2023} developed an autonomous molecular discovery platform to design molecules that simultaneously optimize absorption wavelength, lipophilicity, and photo-oxidative stability. A graph completion model---a type of generative model for molecular design---was used to generate a set of candidate molecules. Candidates for which ASKCOS was able to find a route were prioritized based on a set of expert-curated rules \cite{koscher_autonomous_2023}. \citet{koscher_autonomous_2023} provided us with a set of 121 candidate molecules from one design cycle that passed the retrosynthesis filter. Each candidate was scored according to property prediction models and ranked from 1 to 14 with non-dominated sorting. Non-dominated ranks were converted to rewards between 0 and 1 according to $\reward = (14 - \text{rank})/13$.  

\begin{figure}
    \centering
    \includegraphics{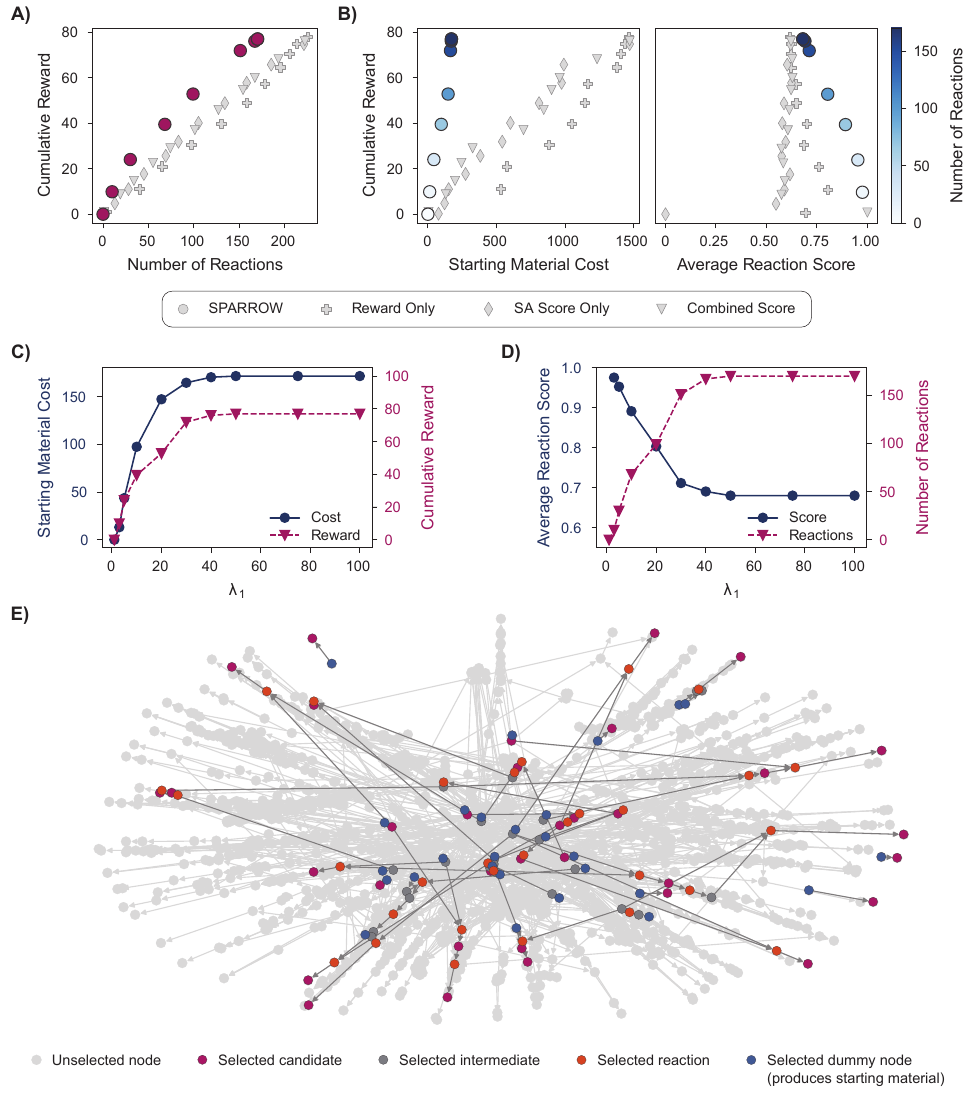}
    \caption{Results of SPARROW applied to an autonomous molecular design cycle by \citet{koscher_autonomous_2023}. (A, B) SPARROW identifies solutions with higher rewards and lower costs when compared to baselines (Tables \ref{S-tab:case-2}, \ref{S-tab:case-2-baselines}). (C)~Increasing the reward weight $\lambda_1$ increases both the cumulative reward and the cost of starting materials. (D) Larger $\lambda_1$ values reduce the relative penalty associated with reaction costs, leading to solutions with more reactions and reactions with lower confidence scores. (E)~SPARROW's downselection can be visualized as a network of selected and unselected nodes ($\lambda_1=5$). All examples use $\lambda_2=\lambda_3=1$. Note that the reported starting material cost is the sum of the price per gram of selected starting materials without consideration of the amount needed. }
    \label{fig:case-2a}
\end{figure}

\begin{figure}
    \centering
    \includegraphics{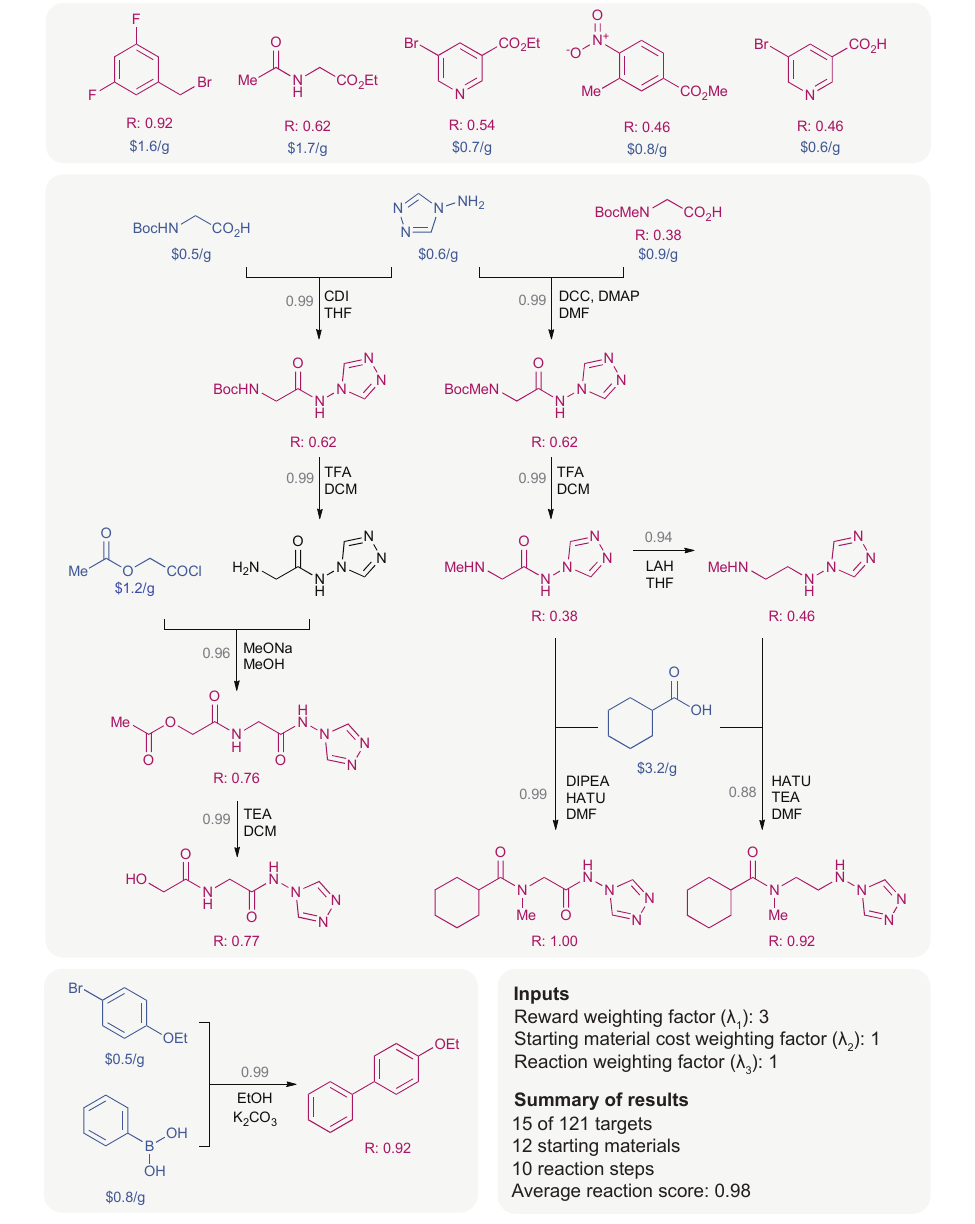}
    \caption{SPARROW's proposed routes for Case 2 with $\boldsymbol{\lambda}=[3,1,1]$. SPARROW selects routes with overlapping reaction steps and intermediates that provide utility themselves. The balance of cost and rewards inherently enables SPARROW to simultaneously propose synthesizing some candidates and buying others that are commercially available. \mollegend}
    \label{fig:case-2b}
\end{figure}

SPARROW discovers distinct sets of batch-efficient routes depending on the values of $\lambda_1$ (Figure \ref{fig:case-2a}A-D). When compared to all baseline approaches, SPARROW selects routes that are comprised of fewer reactions, reactions with greater model confidence scores, and cheaper starting materials (Figure \ref{fig:case-2a}A, B). In Figure \ref{fig:case-2a}E, we depict one of SPARROW's solutions as a downselection from the retrosynthetic graph of all candidate compound and reaction nodes. This candidate set contains buyable compounds, and in some cases, SPARROW proposes directly buying them (Figure \ref{fig:case-2b}). This exemplifies the algorithm's ability to weigh the value of buying candidates and synthesizing others. Although all candidate molecules in this case were proposed by a generative model, the proposed routes demonstrate SPARROW's ability to consolidate molecules proposed by library-based and de novo design. The synthetic routes shown in Figure \ref{fig:case-2b} contain multiple candidate molecules as intermediates. In this way, SPARROW formalizes the common medicinal chemistry mantra of ``test your intermediates'' \cite{barry_lessons_2011, wesolowski_strategies_2016, brown_analysis_2016}. 

\newgeometry{top=1.5cm,bottom=1.5cm, left=1.5cm, right=1.5cm} 

\restoregeometry

\subsection*{Optimizing over large candidate sets}

The third case highlights SPARROW's ability to optimize routes for candidate sets with hundreds of molecules. This candidate set was defined as 300 alectinib analogs proposed by a reaction rule-based generative model from \citet{button_automated_2019}. Candidates were ranked from 1 to 17 based on their similarity to alectinib, and rewards were set according to $\reward=(17-\text{rank})/16$. A candidate set generated by a reaction-based model is predisposed to be highly synthesizable, so we expected ASKCOS to identify routes to most. ASKCOS identified at least one route to 215 of the 300 candidates with an expansion time of 60s; differences in reaction template and starting material definitions account for the difference of 85. 

\begin{figure}
    \centering
    \includegraphics{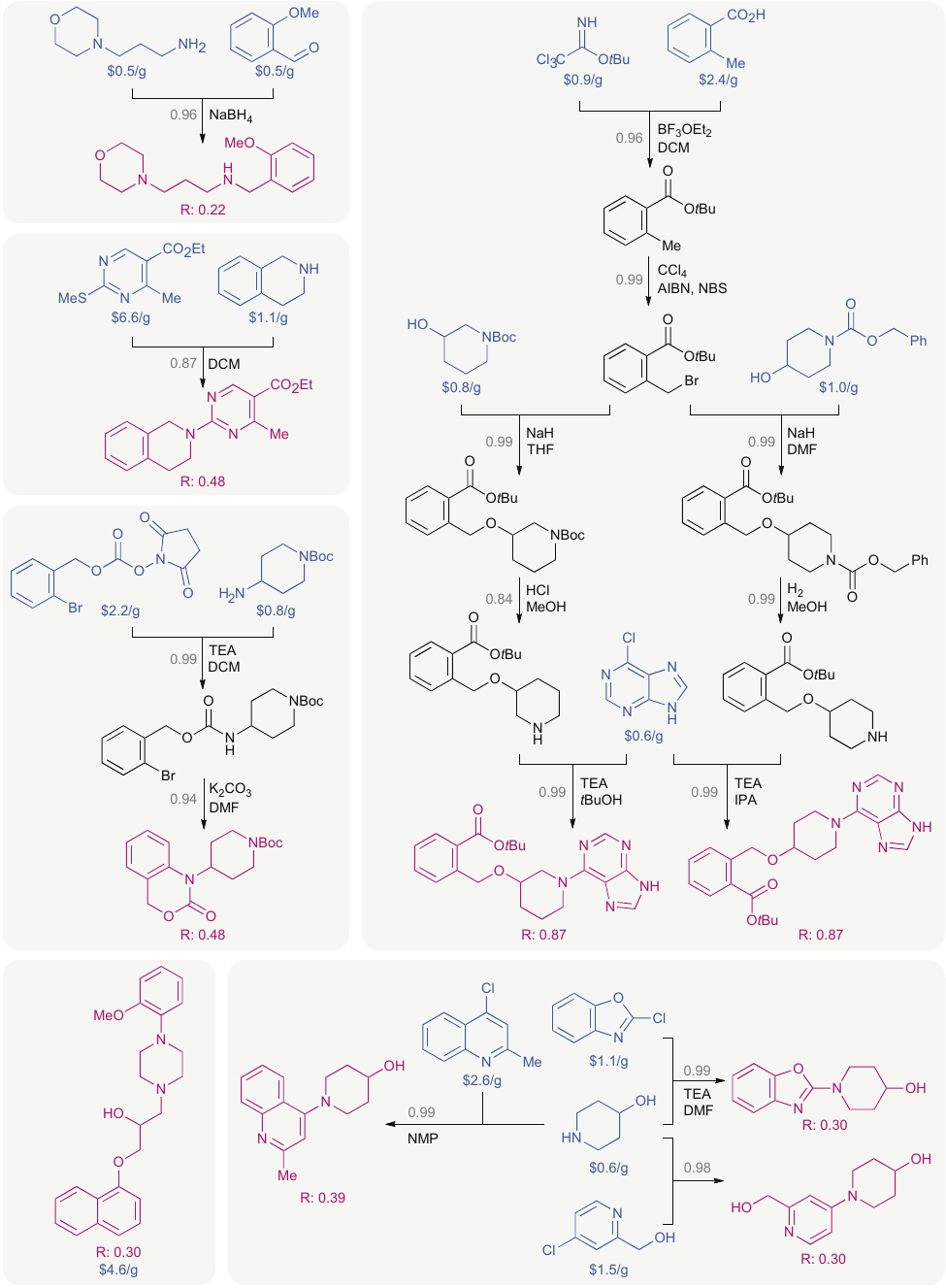}
    \caption{Example set of synthetic routes selected by SPARROW for Case 3 using $\lambda=[30,1,5]$. Synthetic routes are grouped by shared starting materials. SPARROW illustrates that we we may tolerate longer synthetic routes to candidates with higher rewards, demonstrating its ability to balance cost and reward. Common starting materials and commercially available candidates are used where possible. \mollegend}
    \label{fig:case-3b}
\end{figure}

One set of synthetic routes proposed by SPARROW is depicted in Figure \ref{fig:case-3b}. Consistent with the results of previous case studies, SPARROW identifies overlapping reactions steps and starting materials. The two longest of the proposed synthetic routes, presumed to be the most expensive to perform, produce molecules with high rewards. 

For this example of planning a design cycle from a starting pool of 300 candidate molecules, the total runtime of the SPARROW workflow was approximately 13 hours. Approximately 5 hours of retrosynthesis planning, 4 hours of searching for buyability and cost, and 4 hours of condition recommendation and scoring contributed to this computation cost. The optimization problem itself can generally be solved within seconds using PuLP \cite{dunning_pulp_2011} with the open source coin-or branch and cut solver \cite{forrest_coin-orcbc_2023}. Exceptions to the speed of PuLP include some edge cases where weighting factor sets span varying orders of magnitude (e.g., $\boldsymbol{\lambda} = [20, 0.1, 0.1]$)  
and require minutes or hours. This observation aligns with prior findings for a route selection task \cite{gao_direct_2021} and may be caused by numerical instability stemming from arithmetic operations on numbers of different orders of magnitude \cite{klotz_practical_2013, klotz_identification_2014}. Such cases may benefit from decomposition schemes \cite{benders_partitioning_1962}.

Steps preceding the linear optimization---retrosynthetic planning, buyability search, condition recommendation, and reaction scoring---currently contribute most to the time demand of SPARROW. Applying SPARROW to larger candidate sets of thousands or more compounds will be supported by faster retrosynthetic tree search algorithms and faster evaluation of proposed reactions. Improvements in the speed of these steps will facilitate future investigation into the scaling of SPARROW's linear optimization problem to larger candidate sets. 

\newgeometry{top=0.5cm,bottom=1.5cm, left=1.5cm, right=1.5cm} 
\restoregeometry

\section*{Discussion}

The prioritization of molecules for synthesis and testing is fundamentally a balance of cost and utility. Existing methods for molecular design and prioritization are limited in their ability to accurately and simultaneously capture utility and synthetic cost. We have developed SPARROW, an algorithm that prioritizes molecules for synthesis in molecular design cycles. The formulated optimization problem aims to maximize expected information gain while minimizing the cost of synthesis. SPARROW is open source and can accommodate both model-based and chemist-defined synthetic routes. SPARROW is a centralized framework for comparing molecules from various sources and can expand the adoption of existing molecular design tools that inadequately capture synthetic cost. 

The functionality of SPARROW is demonstrated through three case studies. By balancing information gain and cost, SPARROW grasps the impact of batch effects on synthetic cost (i.e., through shared and/or tested intermediates), integrates library-based and de novo design, and deprioritizes structures that are costly or impractical to synthesize. Adjustable weighting factors enable SPARROW to provide a variety of potential batches of molecules and synthetic routes that distinctly balance synthetic cost and utility. 

Future development of SPARROW will relax the assumptions made currently related to synthetic cost and the probability of successful synthesis with a nonlinear objective function (Figure \ref{fig:formulation}). SPARROW's current formulation assumes that molecular utilities are independent, but this is insufficient when optimizing for the diversity of selected candidates or prioritizing matched molecular pairs to uncover subtle structure-activity trends. Additionally, our method currently assumes that reaction costs are constant and independent of other selected reactions. This helps prioritize the use of common intermediates shared by multiple targets but neglects other complexities.\cite{grzybowski_network_2023, wen_chemical_2023} Incorporating the cost of reagents and encouraging selection of reactions that are compatible with parallel, high-throughput, and/or automated synthesis will allow SPARROW to more accurately minimize synthetic cost. Treating synthetic cost as a constrained value instead of a minimized objective may better represent specific projects with well-defined budgets and/or alleviate the challenge of tuning multiple weighting factors. Future work investigating the scaling of SPARROW's linear optimization problem to larger candidate sets will further support its adoption in practical molecular discovery settings. SPARROW's development and expansion to large candidate libraries will also benefit from improvements in enumerative molecular design \cite{levin_computeraided_2023}, retrosynthetic modeling and prediction of reaction success \cite{gotz_high-throughput_2023, casetti_combining_2023}, and molecular property prediction and uncertainty quantification. 

\section*{Methods}

\subsubsection*{Definition of candidate design space}

SPARROW optimizes over a retrosynthetic graph that contains information about candidate molecules and reactions that produce them. The retrosynthetic graph is a directed bipartite graph composed of a set of reaction nodes and a set of compound nodes \cite{pasquini_linchemin_2023, pasquini_linchemin_2024}. The parents of a reaction node are its reactants, and children are reaction products. Parents of a compound node are the reactions that produce that compound as a product, and children of a compound node are reactions that consume it. We store these graphs in \texttt{json} files containing candidate reactions that can be compiled manually with chemist-curated routes or automatically with retrosynthesis models. Candidate reactions do not need to be organized into synthetic routes, as the selection of valid pathways is handled within SPARROW's optimization constraints.    

The balance of cumulative reward, synthetic cost, and the likelihood of reaction success requires additional information: (1) compound buyability and cost, (2) a measure of the likelihood of reaction success, and (3) reaction conditions. We determine the buyability of each compound and its cost, if applicable, using the ChemSpace API \cite{noauthor_chemspace_nodate}. ``Dummy'' parent reaction nodes producing each buyable compound are added to the graph, as implemented by \citet{gao_direct_2021}. SPARROW calls the ASKCOS context recommender \cite{gao_using_2018} to propose reaction conditions and the ASKCOS forward predictor model to estimate the probability of reaction success for each reaction\cite{coley_graph-convolutional_2019}. Reaction conditions are not required by the SPARROW algorithm and are used in our examples only to improve, in principle, the accuracy of the forward prediction model. SPARROW is designed to flexibly accommodate any retrosynthesis, condition recommendation, or reaction scoring model. 

\subsubsection*{Definition of decision variables and constraints}

SPARROW optimizes over two sets of binary decision variables that indicate whether each compound node and each reaction node are included in any of the selected synthetic routes (Figure \ref{fig:formulation}). Three constraints, defined mathematically in Figure \ref{fig:formulation}, ensure that optimized decision variables correspond to valid routes that begin with appropriate starting materials, e.g., commercially available compounds. The first ensures that all parent nodes of a selected reaction node are also selected, as all reactants are required to perform a reaction. The second ensures that if a compound is selected, at least one reaction that produces it must also be selected. For buyable materials deemed worth buying by SPARROW, the corresponding dummy parent node is selected, fulfilling this constraint. This allows the algorithm to choose between buying a molecule and synthesizing it from cheaper starting materials when both options exist. The third constraint forbids the selection of synthetic cycles in the retrosynthetic graph that do not originate from buyable compounds. 

\subsubsection*{Definition of linear objective function}

An ideal problem formulation would perfectly capture the expected information gain per unit cost (Figure \ref{fig:formulation}). However, this objective function is nonlinear with respect to the decision variables. Nonlinear optimization problems provide no guarantee of convergence to a global optimum and require more extensive computational resources to solve. We therefore define a scalarized linear objective function that maximizes cumulative rewards while minimizing synthetic cost and risk of reaction failure  (Figure \ref{fig:formulation}). Both objective functions benefit from solutions that have candidates with high rewards, reactions with high scores, few starting materials, and cheap starting materials. However, because the linear objective is not mathematically equivalent to the nonlinear one, we would not expect the solutions from the two objective functions to be identical in all cases. Weighting factors assign relative importance to each objective. We opt for a scalarized objective function over Pareto optimization in this case because only one solution proposed by SPARROW will ultimately be used. To understand the tradeoff between these competing objectives, SPARROW can be run multiple times with different weighting factors to approximate a Pareto front.  

The first objective maximizes the sum of the rewards of selected candidate molecules. This formulation assumes independent rewards; SPARROW currently does not consider marginal information gain related to molecular diversity and matched molecular pairs. 

The second objective minimizes the sum of all starting materials that must be bought to perform the selected reactions. Costs related to solvents, reagents, and catalysts are not incorporated into SPARROW at this time. Because we account for the commercial availability of starting materials through dummy reaction nodes, this cost is formalized as a cost-weighted sum of decision variables corresponding to the selection of those reactions.

The final objective minimizes penalties associated with selected reactions that are inversely proportional to the calculated probabilities of reaction success. This implicitly minimizes both the total number of selected reaction steps and the risk of reaction failures. Importantly, this objective function assumes that reaction costs are constant and independent of other selected reactions. 

\subsection*{Optimization solver}
SPARROW defines a linear optimization problem based on the decision variables, constraints, and objective function defined above. The linear problem is solved using PuLP \cite{dunning_pulp_2011} with the open source coin-or branch and cut (CBC) solver \cite{forrest_coin-orcbc_2023}. The results in this work were obtained using a relative tolerance of $10^{-7}$ and an absolute tolerance of $10^{-9}$. 

\subsection*{Baselines}
We compare performance of SPARROW with three selection strategies that do not consider the non-additivity of batch synthetic costs. For all baselines, we performed retrosynthetic searches for each candidate with ASKCOS. Output routes are ranked according to a plausibility score, and we selected the route with the highest plausibility score for which all starting materials were identified as buyable by ChemSpace\cite{noauthor_chemspace_nodate} as of March 2024. Selection was performed according to three metrics: (1) reward, (2) synthetic accessibility (SA) score,\cite{ertl_estimation_2009} and (3) a combined score equal to the difference between reward and SA score. We arrived at a set of potential solutions using each strategy by varying the number of selected candidates. The results are compared with SPARROW's solutions in Figures \ref{fig:case-1} and \ref{fig:case-2a}. The discrepancy in compound costs between October 2023, when the costs for SPARROW runs were obtained, and March 2024 are shown in Figure \ref{S-fig:cost-comparison} and Tables \ref{S-tab:case-1-costs} and \ref{S-tab:case-2-costs}. These discrepancies are not significant enough to artificially bolster SPARROW's performance relative to baselines. 

The total time cost of formulating and solving these baselines as their own optimization problems was approximately 60\% lower than  SPARROW. The time cost of retrosynthesis was identical for baselines and SPARROW, but compound price assignment, condition recommendation, and reaction scoring were limited to a smaller set of potential paths, accounting for the difference in overall time cost. 

\section*{Data availability}
SMILES and rewards used for all case studies can be found at \url{github.com/coleygroup/sparrow/tree/main/examples}. All results can be reproduced using included configuration files in the same repository. 

\section*{Code availability and usage}
All code and retrosynthetic routes from ASKCOS used to generate the described results can be found at \url{github.com/coleygroup/sparrow/tree/main/examples}. Full candidate sets with configuration files are included in this repository both for reproducibility and as examples for use of SPARROW. SPARROW is written in Python to facilitate its broad adoption and continued development. 

Detailed instructions to run SPARROW are included in the Github repository. In summary, SPARROW requires as input a list of candidate SMILES strings and corresponding rewards in a \texttt{csv} format. The software outputs a \texttt{json} file containing the SMILES strings of selected candidates and the retrosynthetic route to each in the form of reaction SMILES strings. Because SPARROW defines compound and reaction nodes based on SMILES strings, the limitations of SMILES string notation also apply to SPARROW. Candidate retrosynthetic routes can be compiled into a \texttt{json} file and provided to SPARROW. Alternatively, SPARROW performs retrosynthetic tree searches,\cite{coley_robotic_2019} reaction condition recommendation,\cite{gao_using_2018} and reaction scoring\cite{coley_prediction_2017} if provided the IP address of a deployed ASKCOS instance. Once the retrosynthetic tree is constructed by SPARROW, it can be reused for subsequent runs without access to ASKCOS. Because the cost of running SPARROW is dominated by retrosynthetic graph construction, weighting factors may subsequently be tuned rapidly to arrive at many solutions. Weighting factors must be positive real numbers but are not otherwise constrained. To achieve solutions with more selected candidates, we recommend increasing $\lambda_1$ (the reward weighting factor). Future development of SPARROW will include automatically outputting a set of solutions to alleviate this burden of manual tuning. For more guidelines for implementing SPARROW, please refer to our Github repository at \url{github.com/coleygroup/sparrow}.

\begin{acknowledgement}
This work was supported by the DARPA Accelerated Molecular Discovery program under contract number HR00111920025 and the Office of Naval Research under grant number N00014-21-1-2195. J.C.F. received additional support from the National Science Foundation Graduate Research Fellowship under Grant No. 2141064. We are grateful to Marco Stenta, Marta Pasquini, Daniel Jimenez, and Tobias Ziegler for participating in valuable discussions that guided the development of SPARROW. We also are grateful to Matthew A. McDonald, Brent Koscher, Richard Canty, and the remaining authors of ref. \citenum{koscher_autonomous_2023} for providing the candidate set for Case~2. Finally, we thank Babak Mahjour and Anji Zhang for providing insight into the validity of reactions and conditions proposed by retrosynthetic software.  

\end{acknowledgement}

\section*{Ethics declarations}
The authors declare no competing interests.

\clearpage
\bibliography{references.bib}

\makeatletter\@input{xx.tex}\makeatother
\end{document}


\renewcommand\thepage{S\arabic{page}}
\setcounter{page}{0}
\setcounter{section}{0}

\clearpage

\clearpage \renewcommand\thepage{S\arabic{page}}
\setcounter{page}{1}
\setcounter{section}{0}

\renewcommand{\thefigure}{S\arabic{figure}}
\setcounter{figure}{0}
\renewcommand{\thetable}{S\arabic{table}}
\setcounter{table}{0}

\clearpage
\section{Figures}
\begin{figure}
    \centering
    \includegraphics{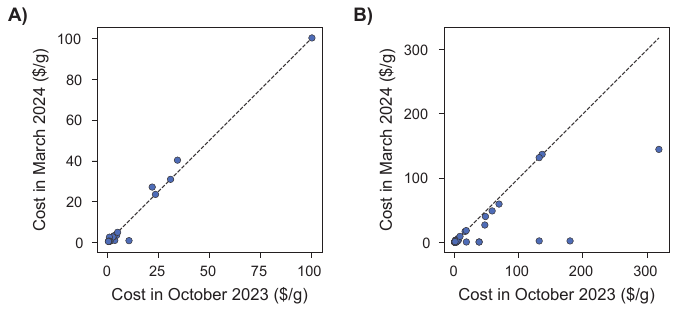}
    \caption{Comparison of compound prices in October 2023 and March 2024. Compound costs used by SPARROW were obtained from the ChemSpace API \cite{noauthor_chemspace_nodate} in October 2023, while baseline calculations were performed with costs obtained in March 2024. We visualize the discrepancy between compound costs between these time periods to confirm that cost discrepancies do not account for SPARROW's performance relative to baselines. (A) Prices for compounds used by both baseline and SPARROW strategies in the first case study. (B) Prices for compounds used by both baseline and SPARROW strategies in the second case study. Cost values are listed in Tables \ref{tab:case-1-costs} and \ref{tab:case-2-costs}.
    }
    \label{fig:cost-comparison}
\end{figure}

\clearpage
\section{Tables}
\begin{table}[H]
\centering
\begin{tabular}{rrrcccc}
\toprule
$\lambda_1\;$ & $\lambda_2\;$ & $\lambda_3\;$ &  \multirow{2}{*}{\shortstack[c]{Cumulative \\ reward}} &  \multirow{2}{*}{\shortstack[c]{Cost of starting \\ materials }} &  \multirow{2}{*}{\shortstack[c]{Number of \\ reaction steps}} &  \multirow{2}{*}{\shortstack[c]{Average reaction \\ score}} \\
&&&&&& \\
\midrule
       2.00 &        1.00 &        1.00 &                 0.0 &                         \phantom{1}0.0 &                         \phantom{1}0 &                     -- \\
       5.00 &        1.00 &        1.00 &                 1.0 &                         \phantom{1}1.3 &                         \phantom{1}3 &                    0.89 \\
       8.00 &        1.00 &        1.00 &                 2.2 &                         \phantom{1}3.8 &                         \phantom{1}8 &                    0.91 \\
      12.50 &        1.00 &        1.00 &                 3.3 &                         \phantom{1}9.3 &                        16 &                    0.96 \\
      15.00 &        1.00 &        1.00 &                 4.3 &                        12.7 &                        24 &                    0.95 \\
      20.00 &        1.00 &        1.00 &                 4.7 &                        13.9 &                        29 &                    0.93 \\
      30.00 &        1.00 &        1.00 &                 5.0 &                        19.1 &                        31 &                    0.93 \\
      50.00 &        1.00 &        1.00 &                 5.6 &                        21.4 &                        39 &                    0.87 \\
      60.00 &        1.00 &        1.00 &                 6.2 &                        23.5 &                        45 &                    0.85 \\
      70.00 &        1.00 &        1.00 &                 6.2 &                        24.3 &                        47 &                    0.84 \\
      80.00 &        1.00 &        1.00 &                 6.2 &                        24.3 &                        47 &                    0.84 \\
            20.00 &        0.00 &        1.00 &                 5.6 &                     14660.7\phantom{100} &                        34 &                    0.96 \\
      20.00 &        0.01 &        1.00 &                 5.2 &                       303.4\phantom{1}&                        28 &                    0.97 \\
      20.00 &        0.02 &        1.00 &                 5.2 &                       109.0\phantom{1}&                        29 &                    0.93 \\
      20.00 &        0.05 &        1.00 &                 5.2 &                        72.7 &                        31 &                    0.96 \\
      20.00 &        0.10 &        1.00 &                 5.2 &                        52.7 &                        32 &                    0.96 \\
      20.00 &        0.20 &        1.00 &                 5.2 &                        22.9 &                        34 &                    0.94 \\
      20.00 &        0.35 &        1.00 &                 5.0 &                        19.1 &                        31 &                    0.93 \\
      20.00 &        0.70 &        1.00 &                 4.7 &                        13.9 &                        29 &                    0.93 \\
      20.00 &        1.30 &        1.00 &                 4.3 &                        12.7 &                        24 &                    0.95 \\
      20.00 &        2.00 &        1.00 &                 4.3 &                        12.7 &                        24 &                    0.95 \\
            20.00 &        1.00 &        0.00 &                 6.2 &                        15.0 &                        72 &                    0.72 \\
      20.00 &        1.00 &        0.10 &                 6.2 &                        18.7 &                        48 &                    0.79 \\
      20.00 &        1.00 &        0.25 &                 6.2 &                        21.3 &                        48 &                    0.85 \\
      20.00 &        1.00 &        0.35 &                 5.6 &                        19.2 &                        42 &                    0.86 \\
      20.00 &        1.00 &        0.50 &                 4.9 &                        14.4 &                        33 &                    0.92 \\
      20.00 &        1.00 &        1.00 &                 4.7 &                        13.9 &                        29 &                    0.93 \\
      20.00 &        1.00 &        1.50 &                 4.3 &                        12.7 &                        24 &                    0.95 \\
      20.00 &        1.00 &        1.70 &                 4.3 &                        12.7 &                        24 &                    0.95 \\
      20.00 &        1.00 &        2.00 &                 3.3 &                         \phantom{1}9.3 &                        16 &                    0.96 \\
      20.00 &        1.00 &        3.00 &                 2.2 &                         \phantom{1}3.8 &                         \phantom{1}8 &                    0.91 \\
      20.00 &        1.00 &        4.00 &                 1.6 &                         \phantom{1}3.2 &                         \phantom{1}5 &                    0.91 \\
      20.00 &        1.00 &        5.00 &                 1.0 &                         \phantom{1}1.3 &                         \phantom{1}3 &                    0.89 \\
\bottomrule
\end{tabular}
    \caption{Summary of SPARROW results when varying $\lambda_1$, $\lambda_2$, and $\lambda_3$ for the first case study \cite{garibsingh_rational_2021}. $\lambda_1$ is the weighting factor associated with candidate rewards, and $\lambda_2$ and $\lambda_3$ correspond to objectives related to starting material costs and reactions, respectively. As shown, weighting factors determine the set of routes proposed by SPARROW. }
    \label{tab:case-1}
\end{table}

\clearpage 

\begin{table}[H]
\centering
\footnotesize
\begin{tabular}{cccccc}
\toprule
Strategy & \multirow{2}{*}{\shortstack[c]{Number of \\ compounds}} &  \multirow{2}{*}{\shortstack[c]{Cumulative \\ reward}} &  \multirow{2}{*}{\shortstack[c]{Cost of starting \\ materials }} &  \multirow{2}{*}{\shortstack[c]{Number of \\ reaction steps}} &  \multirow{2}{*}{\shortstack[c]{Average reaction \\ score}} \\
&&&&& \\
\midrule
   Reward Only &  \phantom{0}1 &                1.0 &                     \phantom{10}9.7 &                    \phantom{0}5 &                    0.71 \\
   Reward Only &  \phantom{0}2 &                1.7 &                    \phantom{0}17.2 &                   10 &                    0.84 \\
   Reward Only &  \phantom{0}3 &                2.3 &                   117.6 &                   12 &                    0.86 \\
   Reward Only &  \phantom{0}4 &                2.9 &                   119.4 &                   15 &                    0.88 \\
   Reward Only &  \phantom{0}5 &                3.5 &                   143.0 &                   18 &                    0.89 \\
   Reward Only &  \phantom{0}6 &                4.0 &                   143.6 &                   20 &                    0.87 \\
   Reward Only &  \phantom{0}7 &                4.5 &                   174.6 &                   25 &                    0.89 \\
   Reward Only &  \phantom{0}8 &                4.9 &                   201.8 &                   29 &                    0.85 \\
   Reward Only &  \phantom{0}9 &                5.3 &                   203.2 &                   34 &                    0.78 \\
   Reward Only & 10 &                5.7 &                   276.9 &                   39 &                    0.75 \\
   Reward Only & 11 &                6.0 &                   281.3 &                   41 &                    0.75 \\
   Reward Only & 12 &                6.2 &                   322.6 &                   44 &                    0.76 \\
   Reward Only & 13 &                6.2 &                   323.8 &                   46 &                    0.76 \\
   Reward Only & 14 &                6.2 &                   331.3 &                   50 &                    0.73 \\
 SA Score Only &  \phantom{0}1 &                0.1 &                     \phantom{10}1.8 &                    \phantom{0}2 &                    0.64 \\
 SA Score Only &  \phantom{0}2 &                0.6 &                    \phantom{0}25.4 &                    \phantom{0}4 &                    0.64 \\
 SA Score Only &  \phantom{0}3 &                0.6 &                    \phantom{0}32.9 &                    \phantom{0}8 &                    0.54 \\
 SA Score Only &  \phantom{0}4 &                1.2 &                    \phantom{0}33.7 &                   11 &                    0.65 \\
 SA Score Only &  \phantom{0}5 &                1.6 &                    \phantom{0}64.1 &                   15 &                    0.64 \\
 SA Score Only &  \phantom{0}6 &                2.3 &                   170.5 &                   18 &                    0.70 \\
 SA Score Only &  \phantom{0}7 &                2.7 &                   204.2 &                   23 &                    0.76 \\
 SA Score Only &  \phantom{0}8 &                3.0 &                   208.6 &                   25 &                    0.76 \\
 SA Score Only &  \phantom{0}9 &                3.6 &                   210.4 &                   28 &                    0.78 \\
 SA Score Only & 10 &                4.3 &                   214.4 &                   33 &                    0.81 \\
 SA Score Only & 11 &                4.5 &                   255.7 &                   36 &                    0.82 \\
 SA Score Only & 12 &                4.9 &                   329.4 &                   41 &                    0.78 \\
 SA Score Only & 13 &                5.9 &                   329.9 &                   45 &                    0.77 \\
 SA Score Only & 14 &                6.2 &                   331.3 &                   50 &                    0.73 \\
Combined Score &  \phantom{0}1 &                0.6 &                    \phantom{0}24.2 &                    \phantom{0}2 &                    0.63 \\
Combined Score &  \phantom{0}2 &                1.6 &                    \phantom{0}33.9 &                    \phantom{0}7 &                    0.69 \\
Combined Score &  \phantom{0}3 &                1.6 &                    \phantom{0}35.1 &                    \phantom{0}9 &                    0.68 \\
Combined Score &  \phantom{0}4 &                2.3 &                   135.5 &                   11 &                    0.73 \\
Combined Score &  \phantom{0}5 &                2.8 &                   136.3 &                   14 &                    0.78 \\
Combined Score &  \phantom{0}6 &                3.5 &                   143.0 &                   19 &                    0.83 \\
Combined Score &  \phantom{0}7 &                4.1 &                   144.8 &                   22 &                    0.84 \\
Combined Score &  \phantom{0}8 &                4.6 &                   175.8 &                   27 &                    0.87 \\
Combined Score &  \phantom{0}9 &                5.0 &                   203.0 &                   31 &                    0.84 \\
Combined Score & 10 &                5.0 &                   210.5 &                   35 &                    0.79 \\
Combined Score & 11 &                5.3 &                   214.9 &                   37 &                    0.79 \\
Combined Score & 12 &                5.7 &                   288.6 &                   42 &                    0.76 \\
Combined Score & 13 &                6.1 &                   290.0 &                   47 &                    0.72 \\
Combined Score & 14 &                6.2 &                   331.3 &                   50 &                    0.73 \\
\bottomrule
\end{tabular}
    \caption{Summary of baseline results for the first case study. SPARROW was compared with downselection strategies that utilize reward values only, synthetic accessibility (SA) scores \cite{ertl_estimation_2009} only, and a combined score equal to the difference between reward and SA score. }
    \label{tab:case-1-baselines}
\end{table}

\clearpage

\begin{table}[h]
    \centering
\begin{tabular}{rrrcccc}
\toprule
$\lambda_1\;$ & $\lambda_2\;$ & $\lambda_3\;$ &  \multirow{2}{*}{\shortstack[c]{Cumulative \\ reward}} &  \multirow{2}{*}{\shortstack[c]{Cost of starting \\ materials }} &  \multirow{2}{*}{\shortstack[c]{Number of \\ reaction steps}} &  \multirow{2}{*}{\shortstack[c]{Average reaction \\ score}} \\
&&&&&& \\
\midrule
       1.00 &        1.00 &        1.00 &                 \phantom{1}0.0 &                         \phantom{10}0.0 &                          \phantom{10}0 &                     -- \\
       3.00 &        1.00 &        1.00 &                  \phantom{1}9.8 &                         \phantom{1}13.1 &                         \phantom{1}10 &                    0.98 \\
       5.00 &        1.00 &        1.00 &                24.0 &                         \phantom{1}43.6 &                         \phantom{1}30 &                    0.95 \\
      10.00 &        1.00 &        1.00 &                39.5 &                         \phantom{1}97.4 &                         \phantom{1}68 &                    0.89 \\
      20.00 &        1.00 &        1.00 &                52.8 &                       147.2 &                         \phantom{1}99 &                    0.80 \\
      30.00 &        1.00 &        1.00 &                71.8 &                       164.5 &                       151 &                    0.71 \\
      40.00 &        1.00 &        1.00 &                76.0 &                       170.3 &                       167 &                    0.69 \\
      50.00 &        1.00 &        1.00 &                76.9 &                       171.5 &                       170 &                    0.68 \\
      75.00 &        1.00 &        1.00 &                76.9 &                       171.5 &                       170 &                    0.68 \\
     100.00 &        1.00 &        1.00 &                76.9 &                       171.5 &                       170 &                    0.68 \\
\bottomrule
\end{tabular}
    \caption{Summary of SPARROW results when varying $\lambda_1$ in the second case study \cite{koscher_autonomous_2023}. $\lambda_1$ is the weighting factor associated with candidate rewards, and $\lambda_2$ and $\lambda_3$ correspond to objectives related to starting material costs and reactions, respectively. As shown, weighting factors determine the set of routes proposed by SPARROW. }
    \label{tab:case-2}
\end{table}

\begin{table}[H]
\centering
\footnotesize
\begin{tabular}{cccccc}
\toprule
Strategy & \multirow{2}{*}{\shortstack[c]{Number of \\ compounds}} &  \multirow{2}{*}{\shortstack[c]{Cumulative \\ reward}} &  \multirow{2}{*}{\shortstack[c]{Cost of starting \\ materials }} &  \multirow{2}{*}{\shortstack[c]{Number of \\ reaction steps}} &  \multirow{2}{*}{\shortstack[c]{Average reaction \\ score}} \\
&&&&& \\
\midrule
   Reward Only &   \phantom{0}1 &                \phantom{0}1.0 &                      \phantom{100}3.0 &                     \phantom{10}4 &                    0.69 \\
   Reward Only &  11 &               11.0 &                    \phantom{0}531.5 &                    \phantom{0}40 &                    0.81 \\
   Reward Only &  21 &               21.0 &                    \phantom{0}575.8 &                    \phantom{0}65 &                    0.76 \\
   Reward Only &  31 &               30.5 &                    \phantom{0}883.6 &                    \phantom{0}97 &                    0.68 \\
   Reward Only &  41 &               39.7 &                  1046.1 &                  130 &                    0.70 \\
   Reward Only &  51 &               48.8 &                  1140.9 &                  158 &                    0.65 \\
   Reward Only &  61 &               57.2 &                  1164.4 &                  178 &                    0.65 \\
   Reward Only &  71 &               64.5 &                  1375.5 &                  195 &                    0.62 \\
   Reward Only &  81 &               70.4 &                  1404.9 &                  205 &                    0.62 \\
   Reward Only &  91 &               74.8 &                  1435.9 &                  213 &                    0.62 \\
   Reward Only & 101\phantom{0} &               77.8 &                  1461.5 &                  225 &                    0.62 \\
 SA Score Only &   \phantom{0}1 &                \phantom{0}0.2 &                     \phantom{10}77.3 &                     \phantom{10}1 &                    0.00 \\
 SA Score Only &  11 &                \phantom{0}4.8 &                    \phantom{0}121.5 &                    \phantom{0}13 &                    0.55 \\
 SA Score Only &  21 &               11.0 &                    \phantom{0}151.0 &                    \phantom{0}28 &                    0.59 \\
 SA Score Only &  31 &               17.7 &                    \phantom{0}274.3 &                    \phantom{0}45 &                    0.62 \\
 SA Score Only &  41 &               25.6 &                    \phantom{0}381.7 &                    \phantom{0}69 &                    0.58 \\
 SA Score Only &  51 &               31.8 &                    \phantom{0}541.6 &                    \phantom{0}83 &                    0.60 \\
 SA Score Only &  61 &               40.1 &                    \phantom{0}601.0 &                  105 &                    0.58 \\
 SA Score Only &  71 &               48.8 &                    \phantom{0}809.7 &                  134 &                    0.59 \\
 SA Score Only &  81 &               57.7 &                   \phantom{0}973.9 &                  158 &                    0.62 \\
 SA Score Only &  91 &               65.7 &                   \phantom{0}990.4 &                  185 &                    0.60 \\
 SA Score Only & 101\phantom{0} &               74.7 &                  1470.1 &                  221 &                    0.62 \\
Combined Score &   \phantom{0}1 &                \phantom{0}0.9 &                      \phantom{100}4.5 &                     \phantom{10}1 &                    1.00 \\
Combined Score &  11 &                \phantom{0}8.7 &                    \phantom{0}130.9 &                    \phantom{0}19 &                    0.57 \\
Combined Score &  21 &               14.6 &                    \phantom{0}197.8 &                    \phantom{0}34 &                    0.59 \\
Combined Score &  31 &               22.5 &                    \phantom{0}244.1 &                    \phantom{0}55 &                    0.58 \\
Combined Score &  41 &               28.9 &                    \phantom{0}325.6 &                    \phantom{0}74 &                    0.57 \\
Combined Score &  51 &               36.9 &                    \phantom{0}696.2 &                  101 &                    0.61 \\
Combined Score &  61 &               45.8 &                    \phantom{0}844.0 &                  127 &                    0.61 \\
Combined Score &  71 &               54.5 &                    \phantom{0}900.7 &                  154 &                    0.62 \\
Combined Score &  81 &               59.8 &                    \phantom{0}960.0 &                  167 &                    0.63 \\
Combined Score &  91 &               68.1 &                  1232.2 &                  193 &                    0.63 \\
Combined Score & 101\phantom{0} &               75.9 &                  1469.5 &                  222 &                    0.62 \\
\bottomrule
\end{tabular}
    \caption{Summary of baseline results for the second case study. SPARROW was compared with downselection strategies that utilize reward values only, synthetic accessibility (SA) scores \cite{ertl_estimation_2009} only, and a combined score equal to the difference between reward and SA score. }
    \label{tab:case-2-baselines}
\end{table}

\clearpage
\begin{table}[H]
\centering
\footnotesize
\begin{tabular}{lcc}
\toprule
SMILES & Cost in October 2023 &  Cost in April 2024 \\
\midrule
                      O=C(Cl)c1ccc(Br)cc1 &   3.3 &   3.3 \\
                               OCc1ccccc1 &   2.7 &   2.7 \\
           O=C(O)c1ccc(NC(=O)c2ccccc2)cc1 &  23.6\phantom{0} &  23.6\phantom{0} \\
                           O=C(O)c1ccccc1 &   1.0 &   1.2 \\
      O=C(O)c1ccc(-c2ccc(C(F)(F)F)cc2)cc1 &  34.4\phantom{0} &  40.4\phantom{0} \\
            O=C(O)c1ccc(C(=O)c2ccccc2)cc1 &   4.5 &   3.5 \\
            O=C(OCc1ccccc1)ON1C(=O)CCC1=O &   1.2 &   1.2 \\
                      O=Cc1ccc(C(=O)O)cc1 &   0.6 &   0.6 \\
                           Cc1ccccc1B(O)O &   0.7 &   0.7 \\
                            OB(O)c1ccccc1 &   0.8 &   0.8 \\
  CC(C)(C)OC(=O)N1C[C@@H](O)C[C@H]1C(=O)O &   1.3 &   0.8 \\
                O=C(O)[C@@H]1C[C@H](O)CN1 &   3.6 &   1.0 \\
                                CC(C)(C)O &   2.7 &   2.7 \\
             O=C(O)c1ccc(-c2ccc(F)cc2)cc1 &  31.0\phantom{0} &  31.0\phantom{0} \\
               O=C(Cl)c1ccc(Oc2ccccc2)cc1 & 100.4\phantom{10} & 100.4\phantom{10} \\
O=C(O)[C@@H]1C[C@@H](O)CN1C(=O)OCc1ccccc1 &   1.0 &   0.9 \\
                         N[C@H](CO)C(=O)O &   0.7 &   0.8 \\
                                   C=CCBr &   3.2 &   3.2 \\
            CC(C)(C)OC(=O)OC(=O)OC(C)(C)C &   2.7 &   2.7 \\
                O=C(Cl)c1ccc(C(F)(F)F)cc1 &   1.0 &   2.6 \\
                        O=C(Cl)OCc1ccccc1 &   5.0 &   5.0 \\
 COC(=O)[C@@H]1C[C@H](O)CN1C(=O)OC(C)(C)C &   1.3 &   0.6 \\
              O=C(Cl)c1ccc(OCc2ccccc2)cc1 &  22.0\phantom{0} &  27.2\phantom{0} \\
           CC(C)(C)OC(=O)N[C@H](CO)C(=O)O &   0.8 &   0.6 \\
                  O=C(O)c1ccccc1-c1ccccc1 &   0.6 &   0.6 \\
COC(=O)[C@@H]1C[C@H](O)CN1C(=O)OCc1ccccc1 &  10.6\phantom{0} &   0.9 \\
                O=C(O)C(c1ccccc1)c1ccccc1 &   0.5 &   0.5 \\
\bottomrule
\end{tabular}
    \caption{Starting material prices from the ChemSpace API \cite{noauthor_chemspace_nodate} in October 2023 and April 2024 used in the first case study, plotted in Figure \ref{fig:cost-comparison}A. }
    \label{tab:case-1-costs}
\end{table}

\begin{table}[H]
\centering
\tiny
\begin{tabular}{lcc}
\toprule
SMILES & Cost in October 2023 &  Cost in April 2024 \\
\midrule
                 O=C1CCC(=O)N1O &   0.5 &   0.5 \\
             O=C(CBr)OCc1ccccc1 &   0.7 &   0.6 \\
            COC(=O)c1cc(Br)ccn1 &   2.6 &   1.7 \\
      CN(CC(=O)O)C(=O)OC(C)(C)C &   0.9 &   0.9 \\
     CC(C)(C)[Si](C)(C)OCC(=O)O & 318.2\phantom{10} & 144.8\phantom{10} \\
         COC(=O)c1ccc(Br)c(C)c1 &   1.3 &   1.4 \\
  Cc1cc(C(=O)O)ccc1[N+](=O)[O-] &   3.2 &   3.2 \\
          CN(CC(=O)O)S(C)(=O)=O &  48.4\phantom{0} &  40.4\phantom{0} \\
           CCOC(=O)c1cc(Br)ccn1 &   6.0 &   6.4 \\
             Cc1cc(C(=O)O)ccc1N &   0.7 &   0.6 \\
                           CNOC &  38.6\phantom{0} &   0.9 \\
                      NC(=O)CCl &   3.2 &   2.5 \\
                           NCCO &   3.2 &   2.4 \\
                      CNCC(N)=O & 180.0\phantom{0} &   2.4 \\
                       O=C(O)CO &   3.2 &   2.2 \\
         Cc1ccc(S(=O)(=O)Cl)cc1 &   3.2 &   3.2 \\
      CC(C)(C)OC(=O)NS(C)(=O)=O &  16.7\phantom{0} &  17.1 \\
                  Brc1cncc(I)c1 &   4.7 &   4.1 \\
          CS(=O)(=O)OS(C)(=O)=O &   0.6 &   0.6 \\
                    CCOC(=O)CNC & 132.0\phantom{10} &   2.4 \\
CC(C)(C)OC(=O)ON=C(C\#N)c1ccccc1 &   1.5 &   3.2 \\
                CC(C)[C@H](N)CO &   1.0 &   1.0 \\
         O=C(O)c1cnc(C(=O)O)cn1 &   2.5 &   1.5 \\
                 O=C(O)C(F)(F)F &   2.7 &   0.5 \\
             CCOc1ccc(B(O)O)cc1 &   1.2 &   1.3 \\
               CC(C)(C)OCC(=O)O &   6.1 &   3.0 \\
         Clc1nc(Cl)c2nc[nH]c2n1 &   1.2 &   1.2 \\
             O=C(Cl)COCc1ccccc1 &   3.4 &   3.9 \\
               CCOC(=O)CNC(C)=O &   1.7 &   1.4 \\
                     Brc1ccccc1 &   3.2 &   3.2 \\
              O=C(O)COCc1ccccc1 &   0.8 &   0.8 \\
                      C1=COCCC1 &   3.2 &   3.2 \\
                         CC(C)N &   2.7 &   1.1 \\
               Cn1cc(C(=O)O)cn1 &   1.6 &   1.7 \\
              CC(=O)N(C)CC(=O)O &  69.5\phantom{0} &  59.8\phantom{0} \\
  CC(C)(C)OC(=O)OC(=O)OC(C)(C)C &   2.7 &   2.7 \\
                 C\#C[Si](C)(C)C &   0.8 &   1.0 \\
           CCOC(=O)c1cnc(Cl)cn1 &  47.4\phantom{0} &  27.2\phantom{0} \\
                   Brc1c[nH]cn1 &   1.7 &   1.9 \\
                  OB(O)c1ccccc1 &   0.8 &   0.8 \\
        O=C(O)CNC(=O)OCc1ccccc1 &   0.5 &   0.5 \\
                O=C(Cl)C1CCCCC1 &   2.7 &   2.7 \\
                          CNCCO &   3.2 &   3.2 \\
                            CCI &   2.8 &   3.2 \\
             O=C(O)c1cncc(Br)c1 &   0.6 &   0.5 \\
            COc1ccc(C(=O)Cl)cc1 &   2.2 &   2.2 \\
                        CC(O)CN &   3.2 &   1.9 \\
           CC(C)(C)[Si](C)(C)Cl &   2.3 &   3.7 \\
            Cc1cc(C(=O)O)ccc1Br &   1.0 &   0.8 \\
              Cc1cccc(C(=O)O)c1 &   1.2 &   1.2 \\
    O=C(OC(=O)C(F)(F)F)C(F)(F)F &   2.7 &   2.7 \\
                  Nc1cc(Cl)ncn1 &   0.6 &   0.6 \\
                       NCC(=O)O &   0.6 &   0.6 \\
              Cn1cc(C(=O)Cl)cn1 & 136.6\phantom{10} & 137.1\phantom{10} \\
          COc1ccc(C(=O)[O-])cc1 &   3.0 &   3.2 \\
                     CCOC(=O)CN &  38.5\phantom{0} &   0.7 \\
                      CNCC(=O)O &   1.2 &   1.0 \\
             COc1ccc(C(=O)O)cc1 &   3.0 &   3.2 \\
                        CC(=O)O &   3.0 &   0.5 \\
                       CC(=O)Cl &   2.7 &   2.7 \\
                       Nn1cnnc1 &   0.6 &   0.6 \\
                     O=C(Cl)CBr &   2.0 &   2.8 \\
            O=S(=O)(Cl)c1ccccc1 &   3.2 &   2.2 \\
           COc1ccc(C(=O)O)cc1Br &   1.6 &   1.7 \\
           Oc1ccc(-c2ccccc2)cc1 &   1.2 &   1.2 \\
                      COC(=O)CN &  18.7\phantom{0} &   0.8 \\
         CC(C)(C)OC(=O)NCC(=O)O &   0.5 &   0.5 \\
              CC(C)(C)OC(=O)CBr &   3.2 &   3.2 \\
                    CS(N)(=O)=O &   1.0 &   1.0 \\
                  O=c1ccnc[nH]1 &   1.9 &   1.9 \\
                      Ic1cccnc1 &   0.9 &   0.7 \\
 O=C(Cl)Oc1ccc([N+](=O)[O-])cc1 &   1.0 &   1.0 \\
                  CN(N=O)C(N)=O &  18.5\phantom{0} &  18.5\phantom{0} \\
                 O=C(O)c1cccnc1 &   0.5 &   0.5 \\
          Nc1nc(Cl)nc2[nH]cnc12 &   3.2 &   2.7 \\
          CC(C)OB(OC(C)C)OC(C)C &   3.2 &   3.2 \\
                    ClCc1ccccc1 & 132.0\phantom{10} & 132.0\phantom{10} \\
                      CC(C)(C)O &   2.7 &   2.7 \\
             CS(=O)(=O)NCC(=O)O &  58.8\phantom{0} &  49.1\phantom{0} \\
               OCc1cc(F)cc(F)c1 &   1.4 &   1.5 \\
                 Clc1cc(Cl)ncn1 &   1.0 &   0.5 \\
                      O=C(O)CCl &   2.7 &   2.4 \\
                C[C@H](O)C(=O)O &   3.2 &   0.9 \\
                 CNc1cc(Cl)ncn1 &   4.1 &   2.1 \\
                CC(=O)OCC(=O)Cl &   1.2 &   1.4 \\
                  Nc1cnc(Cl)nc1 &   8.8 &   9.5 \\
\bottomrule
\end{tabular}
    \caption{Starting material prices from the ChemSpace API \cite{noauthor_chemspace_nodate} in October 2023 and April 2024 used in the second case study, plotted in Figure \ref{fig:cost-comparison}B. }
    \label{tab:case-2-costs}
\end{table}

\clearpage
\bibliography{references.bib}